\documentclass[twocolumn,prd,aps,showpacs,nofootinbib,floatfix]{revtex4}

\begin{document}
\topmargin -0.3in

\input epsf

\title{Numerical evidence for `multi-scalar stars'}

\author{Scott H. Hawley${}^{1}$}
  \thanks{ E-mail address: shawley\,@\,aei.mpg.de}
\author{Matthew W. Choptuik${}^{2}$}
  \thanks{ E-mail address: choptuik\,@\,physics.ubc.ca}
 
\affiliation{${}^1$Max Planck Institut f\"ur Gravitationsphysik, 
               Albert Einstein Institut, D-14476 Golm, Germany}
\affiliation{${}^2$
         CIAR Cosmology and Gravity Program,
         Department of Physics and Astronomy,
         University of British Columbia, Vancouver, British Columbia,
         Canada V6T 1Z1
}

\begin{abstract}
We present a class of general relativistic soliton-like solutions
composed of multiple minimally coupled, massive, real scalar fields
which interact only through the gravitational field.  We describe
a two-parameter family of solutions we call ``phase-shifted boson
stars'' (parameterized by central density $\rho_0$ and phase
$\delta$), which are obtained by solving the ordinary differential
equations associated with boson stars and then altering the phase
between the real and imaginary parts of the field.  These solutions
are similar to boson stars as well as the oscillating soliton stars
found by Seidel and Suen [E. Seidel and W.~M. Suen, Phys. Rev.
Lett. {\bf 66},  1659  (1991)]; in particular, long-time numerical
evolutions suggest that phase-shifted boson stars are stable.  Our
results indicate that scalar soliton-like solutions are perhaps
more generic than has been previously thought.  \end{abstract}

\vspace{-1.0cm}

\maketitle
\section{Introduction}
\vspace{-0.2cm}
The nature of dark matter in the universe is currently an open
question in physics, with many models being proposed to fill this
gap in our understanding, some of which resort to the use of exotic
matter.  Of interest to us is one class of models composed of
massive scalar fields coupled to the general relativistic gravitational
field, from which compact, star-like solutions can be formed,
solutions which go by the names of ``oscillating soliton stars''
(or ``oscillatons'') \cite{OSS,Lopez} for real fields and ``boson
stars'' \cite{Kaup,RB,Colpi,Jetzer,Mielke,Bizon} for complex fields.
These star-like solutions have received renewed attention recently,
and a substantial body of evidence has been advanced in an effort
to show that these fields may be key players on both galactic
\cite{JWLee, Francisco, Miguel} and cosmological \cite{Matos}
scales.   Boson stars have been suggested as alternatives to
primordial black holes \cite{Mielke2} as well as supermassive black
holes in the centers of galaxies \cite{Torres1}, and their
gravitational lensing properties have been explored \cite{Dabrowski},
further developing the treatment of these solutions as objects of
astrophysical interest.

Apart from the possible astrophysical relevance of these star-like
objects, we find them interesting to study from a mathematical
standpoint as well, for their properties as soliton-like solutions
in the nonlinear dynamics of general relativity.  The ``solution
space'' of general relativity is still largely unexplored, and
these scalar objects comprise simple systems with which to conduct
investigations.  It is from this viewpoint that we will proceed in
this paper.

In 1991, Seidel and Suen \cite{OSS} considered the model of a real
massive scalar field, minimally coupled to the general relativistic
gravitational field, with the additional simplifying assumption of
spherical symmetry.  These authors were interested in the existence
of ``nontopological solitons'' in the model: that is, whether the
equations of motion admitted stable, localized, non-singular
distributions of matter which could be interpreted as ``scalar
stars''.  A theorem due to Rosen~\cite{Rosen} suggested that, should
such solutions exist, they could not be static.  Thus, Seidel and
Suen looked for {\em periodic} configurations by substituting a
particular Fourier ansatz into the equations of motion, and solving
the resulting hierarchy of ODEs via a generalized shooting technique.
The authors found strong evidence that periodic star-like solutions
{\em did} exist, and, via direct numerical simulation, demonstrated
that their ``oscillating soliton stars'', if not absolutely stable,
had lifetimes many orders of magnitude longer than the stars'
intrinsic dynamical times.\footnote{More precisely, the oscillating
stars constitute a one-parameter family which may be parametrized
by the mean (period-averaged) central density, $\rho_0$.  As with
other relativistic stellar models, a plot of total (ADM) mass versus
$\rho_0$ exhibits a maximum at $\rho_0^\star$, which seems to
coincide with a transition from stable to unstable configurations.
As expected, only stars with $\rho_0 < \rho_0^\star$ could be stably
evolved for long times.}

These results were surprising to some researchers, particularly since
the model has no conserved Noether current, which, it had been argued,
was responsible for the existence of solitonic solutions in other
non-linear field theories involving scalar fields~\cite{Coleman,Lee}.
However, at least heuristically, we can understand the existence of
the oscillating stars as arising from a balance between the attractive
gravitational interaction and the effective repulsive self-interaction
generated by the mass of the scalar field ({\it i.e.} via the dispersion
relation of the Klein-Gordon equation).

Recently it was shown by Ure\~na-L\'opez \cite{Lopez} that approximate
solutions for boson stars and oscillating soliton stars, or
``oscillatons'' as he calls them, can both be derived from a single
set of equations in a sort of ``stationary limit''.  The similarities
seen between boson stars and oscillating soliton stars in terms of
their curves relating mass, radius and central density can thus be
related formally.

In this paper, we build on the works of Seidel and Suen and Ure\~na-L\'opez 
by considering a matter content consisting of {\em multiple}
scalar fields, and we find some further ways in which boson stars and
oscillating soliton stars are similar.
  For the specific case of two scalar fields, we find
evidence for a new family of quasi-periodic, solitonic 
configurations.\footnote{Although the oscillations appear to be periodic, we 
do not have a proof of their periodicity and so we use the term 
``quasi-periodic''
to describe their temporal behavior.  Small departures from strict periodicity
over long times scales are visible in the simulation results; however,
these departures become smaller as we decrease the radial mesh spacing
$\Delta r$, and thus it is plausible that in the limit 
$\Delta r \rightarrow 0$, the solutions are truly periodic.} 
Together with previous results,
this suggests that solitonic solutions are generic to models which couple
massive scalar fields through the Einstein gravitational field.
We should note, however, that we make no attempt to address the
question of whether these multiple scalar fields actually exist in
nature; rather we are interested in their existence as valid 
mathematical solutions
in the Einstein-Klein-Gordon system.

We begin by considering $n$ real, massive Klein-Gordon fields $\phi_i,
\, i = 1, 2, \ldots n$, without additional self-interaction, minimally
coupled to the general relativistic gravitational field.  Specifically,
choosing units such that $c=1$  and $G = 1$, the Lagrangian density for
the coupled system is

\begin{equation} {\cal L} = \sqrt{-g} \left( R - 
     \sum_{i=1}^{n} 
      \left( {\phi_i}^{;a}{\phi_i}_{;a} - m_i^2\phi_i^2 \right)\right)
 \end{equation}
where $g\equiv\det{g_{\mu\nu}}$, $R$ is the Ricci scalar, 
and $m_i$ is the mass of the $i$-th scalar field.

We now restrict attention to spherical symmetry, and adopt 
the ``polar/areal'' coordinate system, so that the metric takes 
the form:
\begin{equation} 
ds^2 = -\alpha^2(t,r)\,dt^2 +  a^2(t,r)\,dr^2 + r^2\,d\Omega^2 
\label{metric}
\end{equation}
The complete evolution of the scalar fields and spacetime can then be
given in terms of a Klein-Gordon equation for each of the $\phi_i$,
and two constraints derived from the Einstein field equations
and the coordinate conditions used to maintain the metric in the
form~(\ref{metric}).  We solve these equations using the same scheme
adopted for the critical phenomena study described in \cite{HawleyChop},
and only briefly review that scheme here.

We define the following auxiliary scalar field variables:

\begin{equation}\Phi_i\equiv\phi_i',\ \ \ \ \ \ \ 
       \ \ \Pi_i\equiv { a\over\alpha}\dot{\phi}_i\end{equation}
where all variables are functions of $t$ and $r$, 
$\dot{}\equiv\partial/\partial t$ and $'\equiv\partial/\partial r$.
The Klein-Gordon equation is written as the following system:
\begin{eqnarray}
 \dot{\Pi}_i &=& {1\over r^2}\left( {r^2\alpha\over a}\Pi_i \right)' 
                  - m_i^2\alpha a\phi_i  \cr
\dot{\Phi}_i &=& \left( {\alpha\over a}\Pi_i \right)'  \cr
 \phi_i(t,r) &=& \phi_i(t,r_{\rm max}) - 
                 \int_{r_{\rm max}}^{r} \, \Phi(t,{\tilde r}) \, d{\tilde r} 
\label{eq:kg}
\end{eqnarray}
where $r=r_{\rm max}$ is the outer boundary of the computational
domain.  The constraint equations are the ``Hamiltonian constraint''

\begin{equation}  a' =  a{1- a^2\over 2r} 
             + 2\pi r a \sum_{i=1}^{n} 
       \left( {\Pi_i}^2 + {\Phi_i}^2 + a^2 m_i^2\phi_i^2 \right),
\label{eq:ham}
\end{equation}
and the ``slicing condition''

\begin{equation}
  \alpha' = \alpha \left( { a^2-1\over r} + { a'\over  a} 
         - 4\pi r a^2\sum_{i=1}^{n}m_i^2\phi_i^2  \right). 
\label{eq:slice} \end{equation}

For diagnostic purposes, we have also found it useful to compute
and monitor the quantities, $M_i(t,r_{\rm max})$, defined by
\begin{equation}
M_i(t,r_{\rm max}) \equiv 4\pi 
	\int_0^{r_{\rm max}} \tilde{r}^2\, \rho_i(t,\tilde{r})\, d\tilde{r},
\label{eq:defMi}
\end{equation}
where 
\begin{equation}
\rho_i(t,r) =\frac{ \Pi_i{}^2+ \Phi_i{}^2 +  {a^2}\phi_i{}^2}{2a^2}.
\end{equation}
\label{eq:defrhoi}
Loosely speaking, we can interpret $M_i(t,r_{\rm max})$ as the total
 contribution 
of field $i$ to the ADM mass of the spacetime.  In particular, so long 
as no matter out-fluxes through $r=r_{\rm max}$, we have 
\begin{equation}
\sum_{i=1}^{n} M_i(t,r_{\rm max}) = \hbox{{\rm const.}}
\label{eq:conssumMi}
\end{equation}

We solve Eqs. (\ref{eq:kg})-(\ref{eq:slice}) subject to the
the boundary conditions $a(t,0) = 1$ (local flatness at the
origin) and $\alpha(t,r_{\rm max}) = 1/a(t,r_{\rm max})$ (so that $t$
measures proper time as $r\to\infty$).  As in \cite{HawleyChop}, we use
the Sommerfeld condition for a {\em massless} field to set the values
$\phi(t,r_{\rm max}), \Phi(t,r_{\rm max})$ and $\Pi(t,r_{\rm max})$.
Since the Sommerfeld condition is not ideal for a massive field, we ran
our simulations with different values of $r_{\rm max}$, testing for any
periodicity or other effect which might be a function of $r_{\rm max}$,
and usually ran with an $r_{\rm max}$ which was large compared to the time
for which we ran the simulation.  Even with smaller $r_{\rm max}$,
we found our results to
be essentially independent of $r_{\rm max}$ and attribute this to the
fact that there is very little scalar radiation emitted from the 
soliton-like objects considered here.
Our results are also essentially independent of the resolution of the
finite differencing algorithm and the Courant-Friedrichs-Levy 
factor, $\Delta t/\Delta r$,
and we confirm that our results converge in a second-order-accurate
manner using independent residual evaluations.

\vspace{-0.4cm}
\section{``Phase-Shifted Boson Stars''}
\vspace{-0.2cm}
We start by considering the case $n=1$, so that our matter content is
a {\em single} scalar field, $\phi(t,r)$.  We note that the Hamiltonian 
constraint (\ref{eq:ham}) and the slicing condition (\ref{eq:slice}) 
are unchanged if we trivially decompose $\phi$ into two
identical fields ({\it i.e.} now choosing $n=2$), namely $\phi_1(t,r)$
and $\phi_2(t,r)$ = $\phi_1(t,r)$ (with $m_2=m_1=1$), such that
\begin{equation}
 \phi = {1\over \sqrt{2}}\left( \phi_1 + \phi_2 \right).
\label{trivdecomp}
\end{equation}
Further, we note that for fixed $\alpha(t,r)$ and $a(t,r)$, if
$\phi(t,r)$ is a solution of the Klein-Gordon equation (\ref{eq:kg}), then
so is $\kappa\,\phi(t,r)$ where $\kappa$ is an arbitrary real constant.
Since a soliton solution of the system (\ref{eq:kg})-(\ref{eq:slice})
is the oscillating soliton star, we see that a trivial multi-scalar
soliton solution can be obtained by constructing an oscillating soliton
star with a single field $\phi$ (as described in \cite{OSS}) and then
reinterpreting it as a two-field solution
in which $\phi_1 \equiv \phi_2 \equiv \phi/\sqrt{2}$.

Moreover, if we wish to model a boson star
\cite{Kaup,RB,Colpi,Jetzer,Mielke} with no self-interaction potential
(often called a ``mini-boson star'' as in \cite{Bizon}), then we have one
massive complex scalar field $\tilde{\phi}(t,r)$, for which the
real and imaginary parts behave like two real-valued scalar fields:
$\tilde{\phi}(t,r) = \phi_1(t,r) + i\phi_2(t,r)$.  The boson star 
ansatz is $\tilde{\phi}(t,r) = \hat{\phi}(r) \exp(\pm i\omega t),$ 
where $\hat{\phi}(r)$ is real.  This implies
\begin{eqnarray}
  \phi_1(t,r) &=& \hat{\phi}(r)\cos(\omega t) \cr
  \phi_2(t,r) &=& \hat{\phi}(r)\cos(\omega t + \delta),
\label{bsdistrib}
\end{eqnarray}
where $\delta = \mp \pi/2$.

Thus we see that soliton stars and boson stars
can both be obtained using two real scalar fields
with a constant temporal phase difference.
For soliton stars, the fields are identical for all $r$ and $t$; whereas
for boson stars, the fields have identical $r$-dependence, and
the $t$-dependence is the same to within a phase.
(In each case, the central density $\rho_0$ is uniquely fixed by 
the value of the field at the origin, {\it e.g.} $\hat{\phi}(0)$.)

The work described in the remainder of this paper began in the midst of 
our numerical evolutions of
boson stars \cite{HawleyChop}.  The question arose, ``What happens if
we solve for the boson star initial data, then `manually' change the
phase between the two fields, 
and then re-solve the constraints to obtain the metric variables?''
For future reference, we term such a
modified-boson-star configuration a ``phase-shifted boson star.''
This modification to the boson star data
was motivated more by ``practical'' reasons that ``physical'' ones ---
we were interested in studying oscillating soliton stars
but found them difficult to construct.

\begin{figure}
\epsfxsize=8.3cm
\centerline{\epsffile{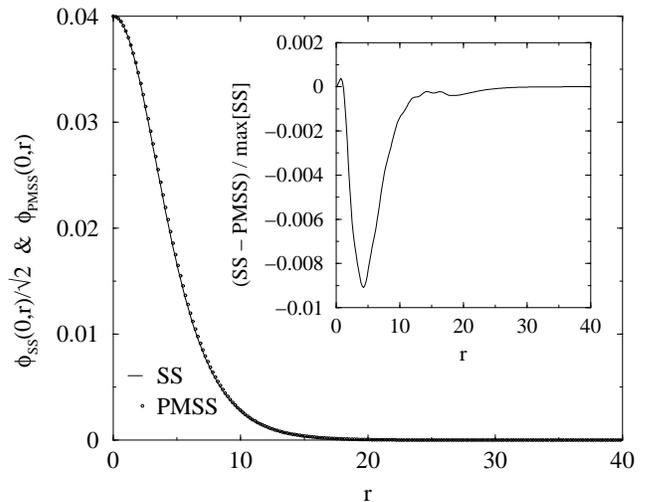}}
\caption{
A comparison between initial data for a true soliton star (SS, solid
line) with the ``poor man's soliton star" (PMSS, circles), which
is a phase-shifted boson star where the two fields are in phase ({\it
i.e.}, $\delta=0$), for a particular choice of the central value of
the field $\phi(0,0)$.   In this figure, we compare the scalar field
of the soliton star with one of the two (identical) fields comprising
the PMSS, where we have divided the soliton star field by $\sqrt{2}$
in keeping with the relation (\ref{trivdecomp}).  
The relative difference between the two solutions is plotted in the 
inset.
Given that these two
solutions are obtained by solving two rather different sets of ODEs
(three simple ODEs for the PMSS, and a 
complicated system of 10 ODEs for the SS), we find it 
remarkable that they are so similar.  
(We note that although we focus on the scalar fields in this figure, the 
metric variables for the PMSS are also close to those of the SS.)
}
\label{fig:pmss_id}
\end{figure}

Taking the boson star initial data and setting $\phi_2(0,r) \equiv
\phi_1(0,r)$ resulted in what might be termed a ``poor man's soliton
star'' (PMSS).  In Figures \ref{fig:pmss_id}, 2 and 3, we show that the
resulting solution is very similar to the true soliton star solution,
and can perhaps best be regarded as a soliton star with a small
perturbation added.

\begin{figure}
\epsfxsize=8.3cm
\vspace{-0.3cm}
\centerline{\epsffile{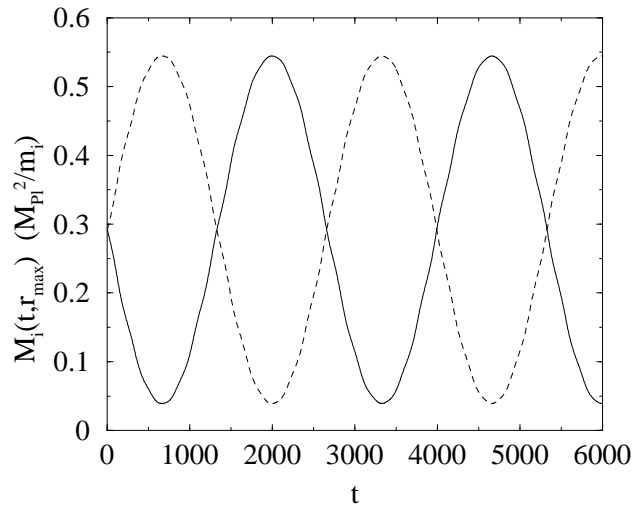}}
\caption{Time development of the quantities $M_1(t,r_{\rm max})$ 
(solid line) and $M_2(t,r_{\rm max})$ (dashed line), defined 
by~(\ref{eq:defMi}), for a phase-shifted boson star with $\delta=-\pi/6$. 
We interpret the behaviour seen in this figure as the periodic transfer of
significant amounts of energy from one scalar field to the other via the
intermediary of the gravitational field.  This is reminiscent of 
``beats'' in weakly-coupled harmonic oscillators; in this case, the 
coupling between the two ``oscillators'' is gravitational.
}
\label{fig:psbs_dm30}
\end{figure}

We then took the boson star initial data $\hat{\phi}(r)$ and distributed
it to $\phi_1$ and $\phi_2$ via Eq. (\ref{bsdistrib}) using some different
value of $\delta$, such as $\delta=-\pi/6$.  (Note that we only apply
Eq. (\ref{bsdistrib}) for the {\em initial data}, {\it i.e.} at $t=0$;
one cannot expect Eq. (\ref{bsdistrib}) to describe the fields for all
$t$ if $\delta\neq \pm \pi/2$, as the $\delta=0$ case demonstrates.)
One aspect of the evolution for such a system can be seen in Figure
\ref{fig:psbs_dm30}.

\begin{figure}
\vspace{-0.25cm}
\epsfxsize=8.3cm
\centerline{\epsffile{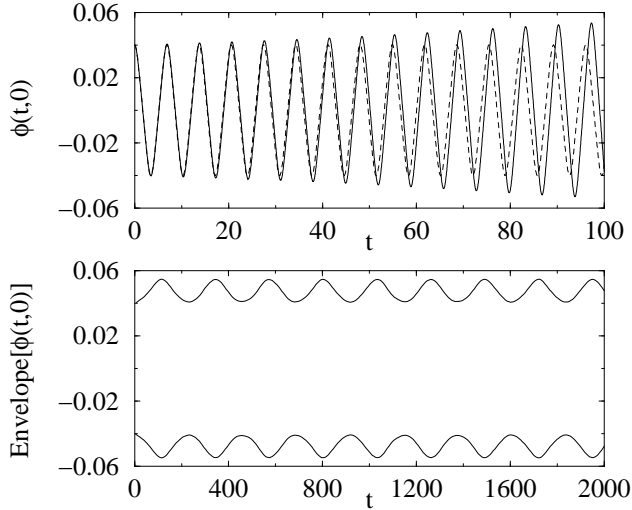}}
\caption
[Central value of the fields $\phi_1(t,0)=\phi_2(t,0)$ vs. time $t$, 
$\delta=0$]
{
Central value of the fields $\phi_1(t,0)=\phi_2(t,0)$ vs. time $t$, for
the ``poor man's soliton star'' (PMSS).
In the top panel we show both the evolution of the PMSS field (solid
line) and, for comparison, a similar evolution for a soliton star field
(dashed line).  In the lower panel we show only the ``envelope'' of the
oscillations in the PMSS scalar field; the period of variations in the envelope
is roughly 32 times the intrinsic period of field.  Stable evolutions
of this system for $t > 20000$ have been obtained.
} \label{fig:pmss_vs_t}
\end{figure}

\begin{figure}
\vspace{-0.3cm}
\centerline{
\epsfxsize=8.3cm
\epsffile{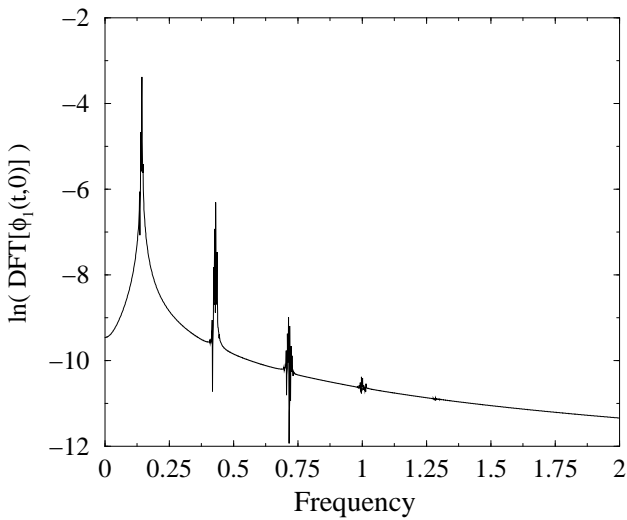}
}
\caption{
Fourier transform of the PMSS evolution shown in Figure
\ref{fig:pmss_vs_t}.  The spikes in the spectrum correspond closely to the
harmonics $\lbrace \omega, 3\omega, 5\omega,...\rbrace$ of the oscillating
soliton star.  The value of $\omega$ is an eigenvalue of the soliton star
ODE problem; for this simulation, $\omega\simeq 0.143.$ } 
\vspace{-0.5cm}
\end{figure}

For each of the many values of $\delta$ we tried, we found an
apparently stable solution which oscillated in some essentially
periodic manner for very long times.  (The phase was preserved
throughout the evolution, {\it i.e.} it is not the case that the
system reverted to a simple ``perturbed boson star'' over time.)
These results led us to conjecture that there may exist a {\it
continuous family} of periodic soliton-like solutions (parameterized
by the phase $\delta$) of which our ``phase-shifted boson stars''
are perturbations.  We hope in the future to construct such a family
{\it directly} via a periodic ansatz of the form used by Seidel
and Suen for their oscillating soliton stars (with additional terms
incorporated to account for the nonlinear coupling between the two
scalar fields).

We would like to mention that, for a given $\hat{\phi}(0)$, varying
$\delta$ has very little effect on the metric variable $a(0,r)$.
Consequently the so-called ``stability curves'' relating total
mass, radius and central density are essentially the same as the
curves for boson stars ($\delta=\pm\pi/2$); consequently, we do
not consider it relevant to plot them here.

\section{Conclusions}
We have strong numerical evidence for the existence of a two-parameter
family of soliton-like solutions to the Einstein-Klein-Gordon system
(parameterized by central density $\rho_0$ and phase $\delta$)
for the case of two scalar fields.  
We speculate that extending the system to
more scalar fields will yield similar results (especially if one uses a
trivial extension like Eq. (\ref{trivdecomp}) ).  The solutions we refer
to as ``phase-shifted boson stars'' consist of boson star initial data
for which the phase difference $\delta$ between the real and imaginary
components of the field have been altered.  These solutions oscillate
in a seemingly periodic manner for very long times; thus they appear to
be stable.  For the case of $\delta=0$, we obtain close approximations
to the oscillating soliton stars of Seidel and Suen.
For other values of $\delta$, we find solutions which also appear to be
stable and periodic; furthermore we can see mass-energy being exchanged
between the two fields.

\acknowledgments
We wish to thank Wai-Mo Suen and Edward Seidel for their valuable
input and for sharing the computer code they used to generate
initial data for oscillating soliton stars.  S.H. Hawley would also
like to thank Alan Rendall, Piotr Bizon and Francisco Guzman for
helpful discussions.  This research was supported by NSERC and
CIAR.  To perform the simulations referred to in this paper, we
used the Silicon Graphics workstations of the Center for Relativity
at the University of Texas at Austin, the CFI-funded {\tt vn} Linux
cluster of the University of British Columbia, and a laptop computer
provided by the Albert Einstein Institute.


\vspace{-0.4cm}
\begin{thebibliography}{99}
\vspace{-0.4cm}
\bibitem{OSS}
E. Seidel and W.~M. Suen, Phys. Rev. Lett. {\bf 66},  1659  (1991).

\bibitem{Lopez}
L. A. Ure\~na-L\'opez, Class.Quant.Grav.19:2617-2632 (2002).

\bibitem{Kaup}
D. Kaup, Phys. Rev. {\bf 172},  1331  (1968).

\bibitem{RB}
R. Ruffini and S. Bonnazzola, Phys. Rev. {\bf 187},  1767  (1969).

\bibitem{Colpi}
M. Colpi, S.~L. Shapiro, and I. Wasserman, Phys. Rev. Lett. {\bf 57},  2485
  (1986).

\bibitem{Jetzer}
P. Jetzer, Phys. Rept. {\bf 220},  163  (1992).

\bibitem{Mielke}
E. W. Mielke and F. E. Schunck, gr-qc/9801063 (1998).

\bibitem{Bizon}
P. Bizo\'n and A. Wasserman, Commun.Math.Phys. {\bf 215} 357 (2000).

\bibitem{JWLee}
J.-W. Lee, I.-G. Koh, Phys.Rev. {\bf D}53, 2236 (1996).

\bibitem{Francisco}
F. S. Guzm\'an and T. Matos, Class.Quant.Grav. 17, L9 (2000).

\bibitem{Miguel}
M. Alcubierre, F. S. Guzm\'an, T. Matos, D. N\'u\~nez, L. A. Ure\~na-L\'opez, 
and P. Wiederhold, Class.Quant.Grav. 19, 5017 (2002).

\bibitem{Matos}
T.~Matos, F.~S.~Guzm\'an and  L. A. Ure\~na-L\'opez, Class.Quant.Grav. 17,
     1707 (2000);
T. Matos and L. A. Ure\~na-L\'opez, Phys.Rev. D{\bf 63}, 063506 (2001).

\bibitem{Mielke2}
E. W. Mielke and  F.~E. Schunck, Nucl.Phys. {\bf B}564, 185 (2000).

\bibitem{Torres1}
D. F. Torres, S. Capozziello, G. Lambiase, Phys.Rev. D{\bf 62}, 104012 (2000).

\bibitem{Dabrowski}
M. P. Dabrowski and F. E. Schunck, astro-ph/9807039 (1998).

\bibitem{Rosen}
G.~Rosen, J.~Math.~Phys. {\bf 7}, 2066 (1966); {\bf 7}, 2071 (1966);
see also
P. Jetzer and D. Scialom, Phys.Lett. {\bf A}169, 12 (1992).

\bibitem{Coleman}
S.~Coleman, Nucl.~Phys. {\bf B262}, 263 (1985).

\bibitem{Lee}
T.~D.~Lee, Phys.~Rev.~D {\bf 35}, 3637 (1987); R.~Friedberg, T.~D.~Lee
and Y.~Pang, Phys.~Rev.~D {\bf 35}, 3658 (1987)

\bibitem{HawleyChop}
S.~H. Hawley and M.~W. Choptuik, Phys. Rev. D{\bf 62},  104024  (2000).



\end{thebibliography}
\end{document}